# CUTTING DOWN ENERGY USAGE IN WIRELESS SENSOR NETWORKS USING DUTY CYCLE TECHNIQUE AND MULTI-HOP ROUTING

Ali Sedighimanesh[1], Mohammad Sedighimanesh[2] and Javad Baqeri[3]

[1,2,3]Department of Electrical, Computer and It Engineering, Islamic Azad University of Qazvin

## Abstract

*A wireless sensor network is composed of many sensor nodes, that have beengiven out in a specific zoneandeach of them hadanability of collecting information from the environment and sending collected data to the sink. The most significant issues in wireless sensor networks, despite the recent progress is the trouble of the severe limitations of energy resources.Since that in different applications of sensor nets, we could throw a static or mobile sink, then all aspects of such networks should be planned with an awareness of energy.One of the most significant topics related to these networks, is routing. One of the most widely used and efficient methods of routing isa hierarchy (based on clustering) method.*

*In The present study with the objective of cutting down energy consumption and persistence of network coverage, we have offered a novel algorithm based on clustering algorithms and multi-hop routing.To achieve this goal, first, we layer the network environment based on the size of the network.We will identify the optimal number of cluster heads and every cluster head based on the mechanism of topology control will start to accept members.Likewise, we set the first layer as gate layer and subsequently identifying the gate's nodes, we'd turn away half of the sensors and then stop using energy and the remaining nodes in this layer will join the gate's nodes because they hold a critical part in bettering the functioning of the system. Cluster heads off following layers send the information to cluster heads in the above layer until sent data will be sent to gate's nodes and finally will be sent to sink. We have tested the proposed algorithm in two situations 1) when the sink is off and 2)when a sink is on and simulation data shows that proposed algorithm has better performance in terms of the life span of a network than LEACH and ELEACH protocols.*

## Keywords:

# 1. INTRODUCTION

One of the most important tools for obtaining information and understanding of the environment which has attracted vast research to itself, is wireless sensor networks. Despite the progress made in this realm of network, sensor nodes due to small size and contingency placement, still depends on disposable batteries for its energy supply. Also usually due to using this type of network in harsh and non-available environments, there is no possibility for recharging or replacing the batteries. Therefore, one of the most important issues in wireless sensor networks is the issue of energy restrictions. Also, since the efficiency of sensor networks is highly depended on life span and coverage of the network, so it's vital to consider energy saving algorithms while designing sensor networks with long life span. Because the sensor nodes work with batteries, they always have a limited life span and recharging for the sensor nodes is always difficult[1]. Operators such as sensing, communication and calculation use most of sensor's energy; transmission of information is considered as the main source of energy consumption[2]. Therefore, theuse of routing algorithms to reduce energy consumption and proper efficiency of bandwidth is vital.

Since nearly all physical phenomena can be sensed with a variety of sensing elements, finding applications in any environment for sensor networks is not surprising. Applications of wireless sensor networks can be divided into two main categories: 1-Viewing or monitoring2-Tracking or Targeting. Other areas of sensor networks are composed of military, environmental, industrial, medicinal and etc [3]–[6]. These applications require sensor networks to be implemented mostly in the form of wireless. Current sensor networks are used on the ground, underwater and underground. A sensor network based on the type of networkenvironment faces with many limitations and challenges. We own a total of five types of sensor networks: mobile wireless sensor networks, multimedia wireless sensor networks, underwater wireless sensor networks and ground wireless sensor networks[7].More often than not, this type of networks has an inactive or very limited dynamic node and a central node which all clients directly (single hop) or indirectly (multi hop) sent their information for it.In direct sending, each sensor sends data directly to the center, therefor due to the great distance between sensors and center, a vast quantity of energy is applied for each transmission. In contrast, designs with shorter communication distances can prolong the lifetime of the network. Therefore, multi-hop connections in this kind of networks are affordable and more efficient than single-hop connections. Scalability is one of the most important factors which can be discussed in wireless sensor networks. Due to limited energy of nodes, the function of all nodes simultaneously causes a drain of energy and reducing the lifetime of nodes. Hence the question of energy is a crisis in the wireless sensor networks by itself. Clustering protocol is an appropriate method for improving the lifetime of wireless sensor networks. In aclustering method, the entire network will be divided into a few clusters. On each cluster, one node is chosen to be Cluster head. Members of clusters will send the processed information to cluster head viasingle-hop or multi-hop routes. Then cluster head send the collected data through single-hop or multi-hop route to sink Fig(1) [8], [9].Clustering approach is an efficient method for adjusting the load between sensor nodes and prolonging the lifetime of the network. With clustering, data sent by the nodes inside the cluster are collected by cluster head or will be sent to sink directly or via intermediary route. So nodes are capable of reducing the communications overload which is resulted from sending the information to sink directly. The purpose of using clustering methods in sensor networks is reducing the volume of sending information and then reducing the required power for communication between nodes. The clustering process creates a two level hierarchy which the upper level consists of cluster head nodes and the lower level consists of sensor nodes which are members of clusters. Cluster head nodes act as a gateway between the sink and sensor nodes.Thesink is the location for processing the information which is received by sensor nodes and also is a place where information will be available to end user. In addition, this structure of sensor nodes and cluster head and sink can be repeated as many times that is necessary and creates multiple layers in a hierarchy wireless sensor network [10].

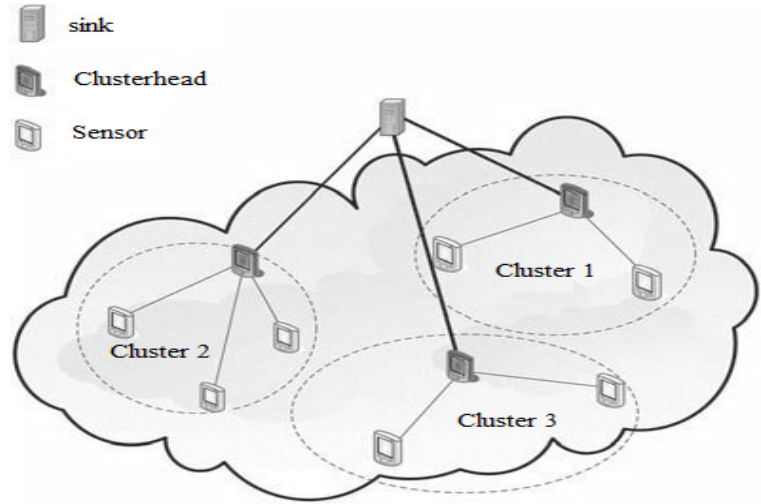

Figure 1. Clustering and data connection in wireless sensor network[10].

## 1.1.Accomplished Works

According to the sensor node resources which are limited to itself and the fact that sensor usually don't have the ability to create a remote connection, so there should be a focus on how to design an effective protocol and its energy to preserve the lifetime of the entire network for special applications of environment. Since the physical layer and the data connection layer are independent of any special software, our attention should be concentrated along the energy aware protocol specifically network layer with energy efficient routing protocols.Yet the purpose of routing protocols in the network layer is dependent on specific programs and limitations of nodes such as energy, memory and computing power.The primary role in the network layers is finding a path for energy efficient routing and helping to transfer approved data from sensor nodes to sink node and also maximizing the lifetime of the network.Scalability as previously remarked, like other telecommunication network, is one of the most significant parameters in designing a wireless sensor network.If all of the network load should be posted on one or some specific routes, volume of traffic on the network will jump significantly with the elongation of the mesh and will dismiss the network efficiency with increased time lag.So, applied that the sensors usually haven't the ability to create a remote connection, the purpose of this approach will limit the expansion of the web. In order to enhance the capability of network coverage in larger regions without causing problems in quality of services, it's recommended to divide the network into several clusters. In this routing approach, every node plays different roles in the network. A clustering algorithm is mainly consisted of two-layer routing in which a layer is used for cluster heads and the other for routing. This algorithm works on a large number of sensor nodes with high density and also it's based on the scalability of routing. The main features of this algorithm are that it divides the receiver wireless network into several clustersusing specific laws. The main purpose of a hierarchical protocol (c.a. Custer-based) is the use of an appropriate method for efficient usage of energy, this is accomplished by utilizing a multi-hop sending as well as the combining of the information of a cluster in order to reduce the amount of sent data. Hierarchical protocols can be split into two groups. Monolayer model, in this model, all nodes play the same role this category includes protocols such as LEACH[11], PEGASIS [12] and modified LEACH algorithms[13], [14]. Hierarchical model, in this algorithm, each network node can start the process of generating a cluster. According to predetermined rules, only one node can do that this category includes protocols such as the TEEN and APTEEN[15][16], HEED[17][18], EEHRP[19], DWEHC [20] and others.

Heinzelman[11], [12] in has proposed a routing algorithm for sensor networks called Hierarchy LEACH. LEACH is one of the most popular hierarchical routing algorithms for sensor networks. LEACH is a clustering protocol that includes distributed information of clusters. LEACH chooses the number of sensors randomly as cluster head (CH) and distributed the energy between them. The idea is that the clustering nodes was done based on the received signal from them and cluster heads was used as routing to the sink. This results in saves energy because instead of all nodes, cluster heads perform transferring operation. Leach is a perfectly distributed and does not need the overall information of the network.However, LEACH uses Single hop routing so that each node can transmit directly to the head and the base station. The optimal number for cluster heads is approximately 5% of the total number of nodes. All processes such as diffusion and aggregation of data is done locally within the cluster. To balance power dissipation of nodes, the cluster heads are randomly changed. We choose a random number r (integer) between 0 and 1. A node in the current period is cluster head if its number is less than the threshold value as follows:

$$T(n) = \begin{cases} \dfrac{p}{1 - p(r.\mathrm{mod}(\dfrac{1}{p}))} & IF \quad n \in G \\ 0 & Otherwise \end{cases} \qquad (1)$$

where p is the percentage of cluster heads and G is a series of nodes that didn't become a cluster in the recent period. Nodes are randomly paired and dynamic clustering increases system lifetime. So it cannot be used for large networks.

HEED [17] is clustering algorithm for wireless sensor networks in which four targets increase network lifetime, clustering phase ending after a certain finite number of iterations, minimizing the control overhead and suitable distribution of clusters are followed by the network. In this protocol, each node as likely (CHprob) proportional to the amount of its remaining energy Decides to be a cluster. This decision is temporary at first and after several iterations is finalized.Nodes that have chosen as cluster head inform this issue to their neighbors. Each of the neighbors, that didn't become a member of the cluster already, now is the cluster member. If the neighbor is already a member of another cluster that its residual energy of its cluster head is lower than the residual energy of new cluster head, this neighbor joins to new cluster head. In addition, if the neighbor is CH itself, after comparing the amount of its residual energy with the residual energy of introducing the cluster, the decision to remain as CH or be transferred to the new cluster. Each cluster, while others decide not to join the cluster, the value of your CHprob doubles again as neighbors introduced a cluster. Each clusterhead, if decide not to join the other cluster, double its value of CHprob introduced itself again as cluster head to its neighbors. If the value of CHprob was greater than one in the node, the node appoints itself as final cluster head. In this case, the neighbors of this node also will be the final clusters member that no change occurred in it. At the end of this phase, if the nodes, didn't receive any cluster introducing message, then decide to be a cluster head itself.

Energy efficient and effective protocols for wireless sensor networks is a necessity to not only reduce total energy consumption in the network, but also to balanced evenly distribution of energy load among the nodes in the network to increase the network lifetime[21].Clustering protocol is a good way to enhance the lifetime of these networks. When the main station is mobile, different circumstances should be considered. We will discussthe proposed method in this article, which includes employing mechanism for layered network environment with respect to the size of the network environment, choosing the closest layer to sink as the gateway layer, how to choose the node for the first layer in order to save energy with respect to having that first layer of coverage is not in trouble, and theclustering and proper selection of  cluster head in each layer separately  and  finally multi-hop sending  the  data  to the  sink.  Given that  the sink  is  in motion and due to the location of the sink, layered environment and transmit data to the base station we should use a variety of mechanisms.To increase lifespan, reduce the cluster head load, balance energy the nodes and reduce delays in data transmission process, we will use different clustering topologies in each hop and level with different methods[22]–[24].We will discuss the phases of the proposed algorithm and then comparing,simulating and conclusion as follows.

## 2.THE PROPOSED ALGORITHM  PHASES

The proposed algorithm is examined in static and dynamic mode sink, in which some parameters and assumptions are the same for two modes, also there are assumptions in whichevery one of them should be considered separately that describing them as follows. In this section and in 5 sections, we will be discussinglayering the network environment, optimum number of cluster heads, clustering model, symmetrizing the number of cluster and routing mode.

### 2.1.Layering The Network Environment

In the proposed method, using the relation (2) the network environment will be layered commensurate with its size and distance to sink.

$$L_{(1)} = (.15) \times Y$$

$$M = Y - L_{(1)}$$

$$L_{(i)} = \begin{cases} IF & M \geq L_{(i-1)} + (.15) \times M \\ Then & L_{(i)} = L_{(i-1)} + (.15) \times M \\ & M = M - L_{(i)} \\ Else & L_{(i-1)} = L_{(i-1)} + M \end{cases} \qquad (2)$$

In relation(2), y is the length of the network, M is length of network minus layers combined, and $L_{(I)}$ is the size of (i) the layer. In the allocation of environment to each layer, the relationship (2) should be checked. Layering mechanism can be started with different values at the beginning, which this creates different values for $L_i$. Finally, with regard to different values, this relationship of $Y = L_1 + L_2 + \ldots + L_i$ should be applied in our layering mechanism.

## 2.2. The Optimal Number Of Cluster Heads

The size of a cluster, the number of cluster heads in each cluster and selection of cluster heads are among the most important factors in clustering. Increasing the number of cluster head cause control information overload and questions of routing. Reducing the number of cluster heads caused the increase in input load and early energy drain in cluster head. In the proposed method, optimal number of cluster heads in every layer is different. After layering the environment, the number of alive nodes in the first layer (N-1) is calculated and then the relationship 3, was used as optimum number of cluster heads in the first layer.

$$K_{opt(1)} = \lceil \%2 \times N_1 \rceil$$

$$IF(K_{opt(1)}) < 2 \, Then \quad K_{opt(1)} = 2 \qquad (3)$$

According to the above formula, if the optimum number of gateway node is a value less than 2, value of 2 will be inserted, i.e. at least 2 gateway nodes must be fitted. in eq. (3)N1 is the number of nodes in the first layer.

$K_{opt(1)}$ determines the optimum number of gateway nodes for first layer. For the next layers, we define every layer such a way that the optimum number of cluster heads increases with increasing the distance to sink in which $K_{opt(i)}$ is calculated through eq. (4):

$$K_{opt(i)} = \left\lceil (\%5 \times N_i) + \frac{1}{2} K_{opt(i-1)} \right\rceil \qquad (4)$$

In the above relations, $N_i$ and $K_{opt(i)}$ are the number of existent nodes and the optimum number of clusters in (i)th layer except first layer, respectively. After selecting the optimum number of the cluster heads, we have different modes for selecting a cluster as a cluster head, in first mode, the same selection structure in LEACH algorithm is used and selecting cluster head is done through a probability function. Each node chooses a random number between 0 and 1 and if the selected number is lower than T(n), then it's selected as current round cluster head.

$$T(n) = \begin{cases} \dfrac{p}{1 - p(r.\mathrm{mod}(\dfrac{1}{p}))} & IF \quad n \in G \\ \\ 0 \quad Otherwise \end{cases} \tag{5}$$

P is the probability of selection as cluster head, r is the number of the current round and G is the set of nodes which haven't been a cluster head in 1/p recent rounds, $E_{in}$ is the primary energy of node and E is the residual energy of a node. In the second mode, we will involve the energy in eq. (5) too i.e the same E-LEACH algorithm will be used.

$$T(n) = \begin{cases} \dfrac{p}{1 - p(r.\mathrm{mod}(\dfrac{1}{p}))} \times \dfrac{E}{E_{in}} & IF \quad n \in G \\ \\ 0 \quad Otherwise \end{cases} \tag{6}$$

In above relationship, $E_{in}$ is the primary energy of node and E is the energy of a node.

## 2.3. Clustering Model

Cluster heads have a vital role in communication between layers in the clustering algorithm. Cluster heads with collecting information from members of each cluster and also from below the layer cluster head and sending them after compression to above layer cluster heads causes sending the information to sink which has been done through a hierarchy of routing. To maintain the energy of cluster heads and maintain the communications like between cluster heads close to sink, it's advisable that the number of members of the clusters near to sink be less tocluster heads near the sink, allocate most of their energy to receive data from lower layers and sending data to upper layers or sink.

## 2.4. Symmetrizing The Size Of Clusters

In the proposed method, clustering in each layer except the first layer is done individually and for each layer after defining the $K_{opt(i)}$, the numbers of nodes which are not cluster heads in every layer is calculated and it's divided to $K_{opt(i)}$ and the maximum number of nodes in every cluster head is calculated through this formula:

$$C_i = \left\lfloor \left\lfloor \dfrac{n_i - K_{opt(i)}}{K_{opt(i)}} \right\rfloor \times (.9) \right\rfloor \tag{7}$$

In the above formula, $n_i$,$K_{opt(i)}$ and $C_i$are denoting the number of alive nodes, the optimum number of cluster heads and the maximum number of members of each cluster in (i)th layer respectively. In this manner, the maximum number of members of cluster head in ith layer equal to Ci. First each cluster head will give Cimembers and then the nodes which haven't joined any cluster heads will join to the nearest cluster head,this approach will reduce wasting energy of nodes. In the first layer, after determination of the number of gateway node, we will need out the amount the first layer minus the number of nodes. And then approximately 50% of nodes which are closer to sink and have more energy considered as dormant nodes to prevent of early discharge of energy in the first layer. Selecting the gateway layers for each round, was done according to the mentioned items in order to select the best nodes to send the received information from the bottom layers to sink.

$$K = \left\lfloor \frac{n_1 - K_{opt(1)}}{2} \right\rfloor \qquad\qquad (8)$$

In relation(8),$n_1$,$K_{opt(1)}$ and K are the number of nodes, the optimum number of cluster heads or the gateway nodes in the first layer and mandated nodes in the current node, respectively. With regard to this fact that more than 50% of the nodes are within the first layer of the network, there will not be any conflict in performance and coverage of the network.

## 2.5.Routing Model

In the proposed method, routing is done between the cluster head from the lower layers to the upper ones. The algorithm operates as cluster head of the lowest layer and higher layers' exchange information. Cluster head in bottom layer will choose the most appropriate cluster head or the best route with regard to remaining energy and distance to upper layer cluster heads and inform itself to the chosen cluster head with a status and request message. This action was done separately for each cluster heads. Algorithm operated similarly for all layers to the first layer which is the closest layer to the sink. In the first layer according to relationship (3),when the optimum number of cluster heads is obtained, half of which are closer to the sink are set too dormant for didn't consume energy and the rest of the nodes will wake up to sense the environment and send their information to gateway nodes which is selected to become a cluster head and is closer to itself. When cluster heads of lower layers receive data from their own members or other lower layer cluster heads, they will aggregate them, and then there will be two modes: First: If there was a gateway node in the first layer, cluster heads of bottom layer send information to the nearest gateway node and data through the gateway node is sent to the sink. Second mode: If there wasn't a gateway node in the first layer (all nodes in the first layer have been eliminated) cluster heads will send their data directly to the sink. The nodes of lower layer may be the best choice nodes and gateway nodes. The bottom layer  nodes can choose the best gateway nodes and all the nodes will communicate with a gateway node with a high probability and this reduces power consumption in the first layer nodes and also not clustering this layer will cause energy storage for nodes because a Phase of cluster formation and recruitment and sending information to cluster head and cluster aggregation is lost by cluster head, the upper layer which is the closest layer to sink to sink will focus on sending the information from the lower layer to sink.

## 2.6. Routing Model in Mobile Sink

When the sink is mobile, layering the environment, calculating the optimal number of cluster heads, model of clustering and symmetrizingcluster size is similar to the static sink mode, only in time of routing, because sink is mobile, conditions of layering should change when the coordinates of sink is changing and in thisway sink should declare its position to nodes wen it's in movement. To do this, we must consider boundaries and when the ink crossed them, layering is also changes proportional to its status. The sink is moved in a jump mode. For example,wespecify in each round that how many meters should a sink move. This is explained in Figure 2.

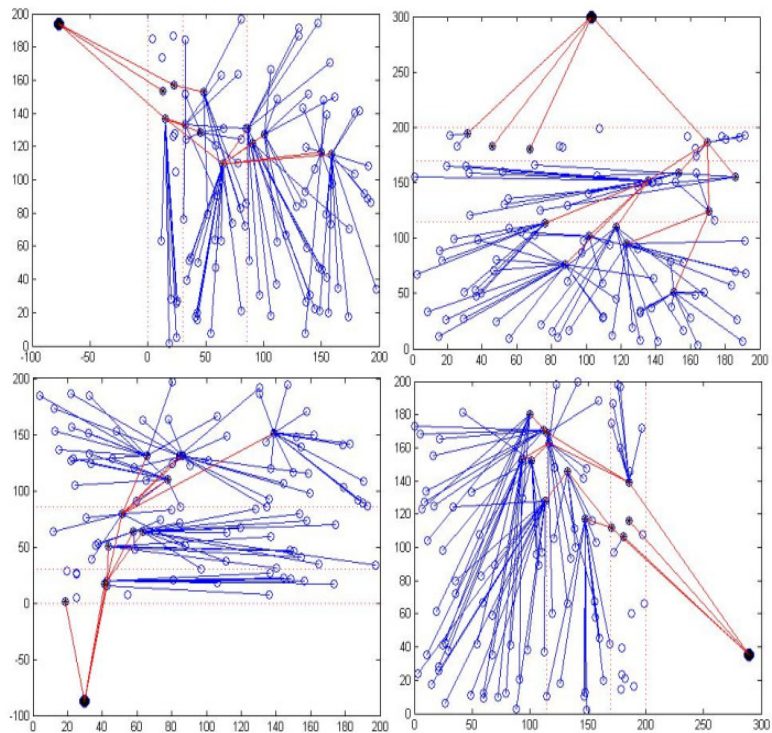

Figure 2. how to move the sink and changes in layeringenvironments.

In this section we will examine theagent of change in the speed sink in lifetime of wireless sensor networks, in order to do so, we will consider assumptions to our stimulation where for this purpose we will consider an environment of $200 \times 200$ by taking 100 sensor network nodes. move speed of sink is considered as a variable of X that is calculated in each round, for example, X = 43 means that sink moves, 43 meters in each round.

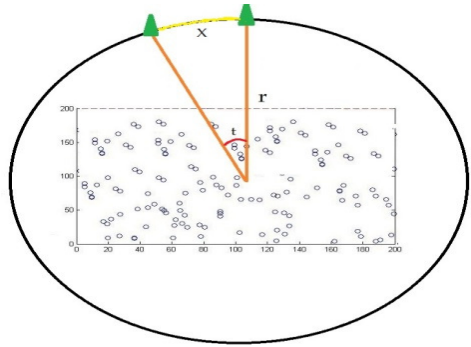

Figure 3. Calculating the distance traveled.

X = t × (pi / 180) × r(9)

  X is the distance traveled in each round, r is the radius of the relevant circle. Pi is the amount equal to 3.14.

## 3. SIMULATION

In this section the results of simulation of the proposed clustering algorithm, analysis and comparing the results with previous clustering protocols will be discussed. The simulation was carried out using MATLAB software. Given that we have a mobile sink, thenat firstwe will consider a basic station, which in each round its position varies based on the speed of the sink. The position of the initial parameters of wireless sensor networks for the simulation of100 nodes is shown in Table 1.

Table 1. wireless sensor network basic parameters.

| Parameter | value |
|---|---|
| Area Size | 200×200 $m^2$ |
| Initial energy in each Node | 0.5 (J) |
| $E_{cpu}$ | 7(nJ/bit) |
| $E_{elec}$ | 50 (nJ/bit) |

| $E_{amp}$ | 0.659 (nJ/m$^2$) |
|---|---|
| Packet Size | 4000Bit |
| The initial position of the sink | 100×300 m$^2$ |

The Following assumptions are considered in the simulation:

• Network environment, is the square of the number of sensors specified.
• Sensors have been released randomly.
• Sensors are aware of their location.
• ID is the unique for each sensor.
• Each node senses the data alternately and always has a data to send.
• The volume of sensed data by node is the same.
• Perform data compression to minimize the amount of data
• Sink is mobile and runs at a constant speed around the environment.
We considered the speed of sink based on each round so that the speed of 10 meters per round is specified.

## 3.1. Energy Model

Energy consumption in wireless sensor network consists of three parts: data, data acquisition and data processing. Energy model is brought in e.q. (10) [22, 24].

$$\begin{cases} P_T(K) = E_{elec} * K + E_{amp} * d^\gamma * K \\ \quad P_R(K) = E_{elec} * k \\ \quad P_{cpu}(K) = E_{cpu} * k \end{cases} \tag{10}$$

PR and $P_{cpu}$ represent the energy consumption of sending, receiving and processing of K bits data. $E_{elec}$c, $E_{amp}$ and $E_{cpu}$ represent the energy consumption (NJ / bit) for sending any bits on the radio radius, the energy to send with a radius of more than $E_{elec}$ and energy per bit is required for processing. According to equation (10) the total energy consumed by k is like in relation(11).

$$P_{Total} = P_{Send} + P_{Receive} + P_{Process}$$
$$P_{کل} = k(2E_{elec} + E_{cpu} + E_{amp} \times d^\gamma) \tag{11}$$

In (11) eq.we see that, energy consumption has a direct relationship with the data length. If the send data isless, we will use less energy. If Transmission distance is less than the threshold, energy consumption will berelated to $d^2$. If the transmission distance is greater than the threshold, it is associated with $d^4$. Therefore, the less the transmission distance is, energy consumption can be reduced.

## 3.2.Simulation Results

In this section simulation, we will compare the results of the proposed algorithms and ELEACH and LEACH algorithms. MATLAB software is used as asimulator. The results are evaluated based on two modes: first mode in static sink and second, in mobile sink, results are described separately as follows.

### 3.2.1.Static Sink

The sink is static in the outfield in this mode.

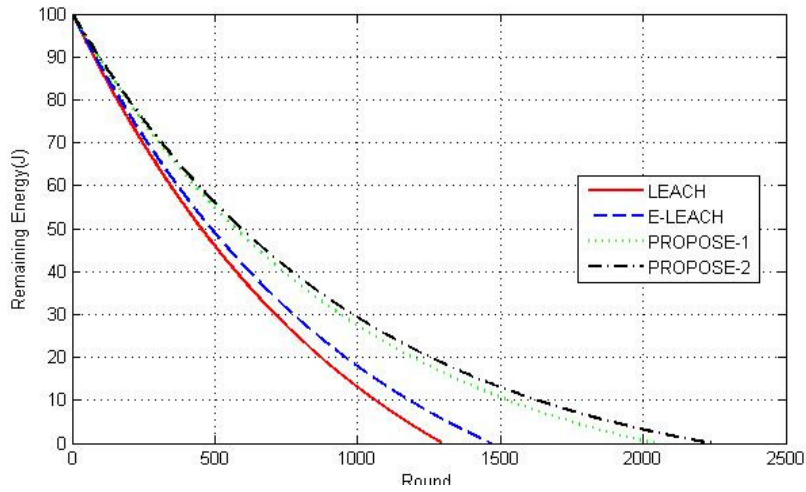

Figure 4. Compare of lifetime between the proposed methods and LEACH and ELEACH protocols.

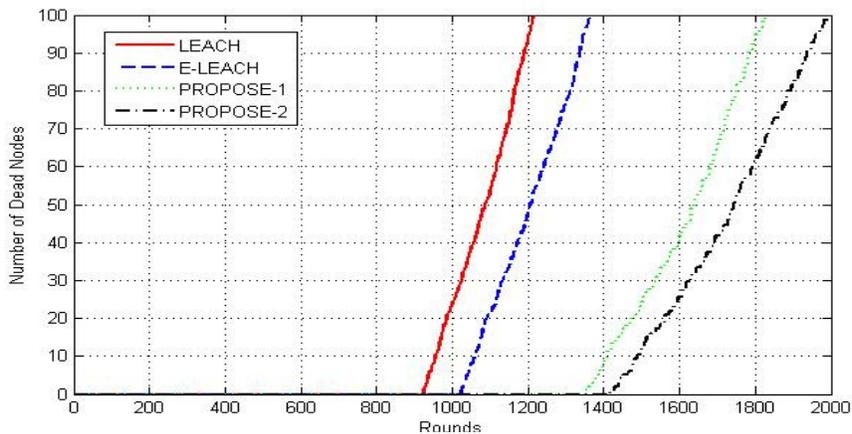

Figure 5. Comparison of number of dead sensor nodes between the proposed methods and LEACH and ELEACH protocols.

As seen in the figure, our proposed methods have improved 49 percent in death of first node than LEACH and 34% than ELEACH and has a 69% improvement in death of last node than LEACH and 34% percent than ELEACH. In Table (2) the impact of increasing the number of nodes of sensor in networks, according to the assumptions in table (1)is discussed.

Table 2. impacts of the number of nodes on networks lifetime in satatic sink.

| The number of nodes | | | | | | |
|---|---|---|---|---|---|---|
| 300 nodes | | 200 nodes | | 100 nodes | | |
| The last node | The last node | The first nodes | The first nodes | The last node | The first nodes | Life Time Algorithm |
| 3985 | 2185 | 3095 | 1684 | 2235 | 1415 | Propose2 |
| 3667 | 2017 | 2890 | 1514 | 2036 | 1345 | Propose1 |
| 2214 | 1489 | 1956 | 1202 | 1466 | 1020 | ELEACH |
| 1963 | 1325 | 1769 | 1096 | 1295 | 919 | LEACH |

We can see in Table (2), with an increase of nodes in the environment network also increases the network lifetime. Because the number of nodes is directly proportional to the network's lifetime, the more numbers of nodes is the longer lifetime of the network.

In the following table the impacts of the size of environment on the network's lifetime were compared and assume that the numbers of nodes in different environment is equal to 100 nodes.

Table 3. impact of the size of environment in network lifetime in static sink.

| The size of the network environment $M^2$ | | | | | | |
|---|---|---|---|---|---|---|
| 300×300 | | 200×200 | | 100×100 | | |
| The last node | The last node | The first nodes | The first nodes | The last node | The first nodes | Life Time Algorithm |
| 940 | 595 | 1450 | 919 | 2235 | 1415 | Propose2 |
| 740 | 523 | 1250 | 849 | 2036 | 1345 | Propose1 |
| 448 | 202 | 682 | 524 | 1466 | 1020 | ELEACH |
| 325 | 102 | 510 | 423 | 1295 | 919 | LEACH |

According to Table 4.3, if the size of the environment become larger and the number of sensor nodes does not change, the network lifetime will be decreased.

### 3.2.2. Mobile sink

The sink in this mode, is moving in a specified direction around the environment.

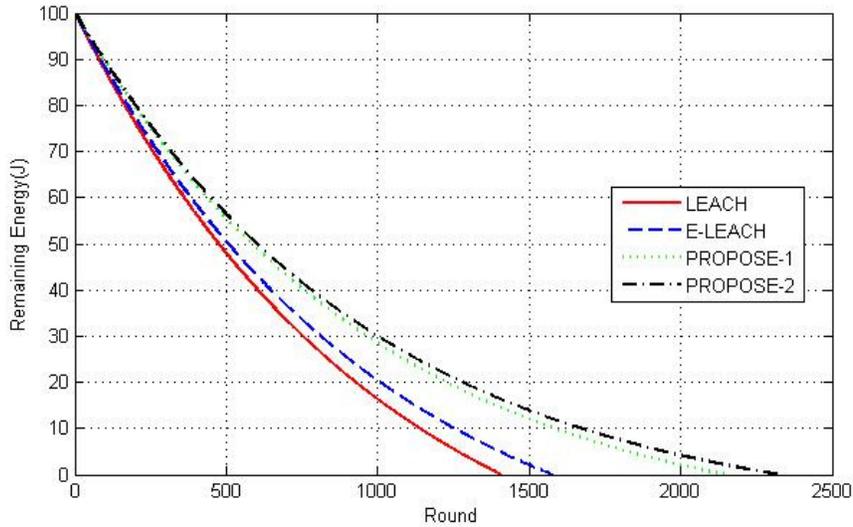

Figure 6. Comparing the lifespan of the proposed methods and LEACH and ELEACH protocols.

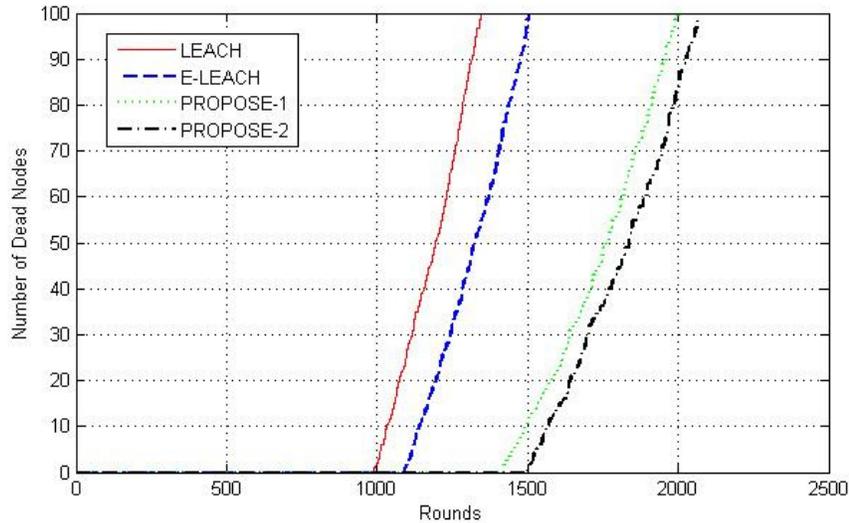

Figure 7. Comparing the number of dead sensor nodes in the proposed methods and LEACH and ELEACH protocols.

As seen in the figure, in proposed methods compared with other protocols, there has been an improvement by 51 in the death of the first nodethan the LEACH and 38% in the death of the last node than the ELEACH and 74% than the LEACH and 52% than the ELEACH. The impact of increasing the number of sensors nodes on networks, according to assumptions in Table (4) was investigated in table (1).

Table 4. Effects of number of nodes in network lifetime in mobile sink.

| The number of nodes | | | | | | |
|---|---|---|---|---|---|---|
| 300 nodes | | 200 nodes | | 100 nodes | | Life Time / Algorithm |
| The last node | The last node | The first nodes | The first nodes | The last node | The first nodes | |
| 4165 | 2278 | 3185 | 1745 | 2325 | 1495 | Propose2 |
| 3860 | 2123 | 3040 | 1595 | 2145 | 1413 | Propose1 |
| 2410 | 1595 | 2105 | 1283 | 1575 | 1088 | ELEACH |
| 2150 | 1431 | 1919 | 1117 | 1405 | 987 | LEACH |

by observing table (4) we found that triggering a sink in this case results in increase the lifetime of the network. In the case of mobile sink, our proposed methods have a better performance than traditional methods, here, the increase in the number of nodes will increase the network lifetime. The following table compares the impact of changes in the size of environment of network on the lifetime of the network, like static sink. It is assumed that the number of nodes in different environment is equal to 100 nodes.

Table 5. The influence of the size of environment on the network lifetime in mobile sink.

| The size of the network environment $M^2$ | | | | | | |
|---|---|---|---|---|---|---|
| 300×300 | | 200×200 | | 100×100 | | Life Time / Algorithm |
| The last node | The last node | The first nodes | The first nodes | The last node | The first nodes | |
| 980 | 630 | 1510 | 970 | 2325 | 1495 | Propose2 |
| 801 | 545 | 1330 | 885 | 2145 | 1413 | Propose1 |
| 575 | 225 | 760 | 55 | 1575 | 1088 | ELEACH |
| 415 | 120 | 590 | 460 | 1405 | 987 | LEACH |

According to Table (3) and (5), the size ofthe network environmentis correlated with the lifetime of the network, the larger the size of the larger environment, the decreased lifetime of network.

### 3.2.3. Change Of Pace In Sinks

This section examines the impact of sink's speed on lifetime of networks. We will set the sink speed in three modes, assuming that the size of simulated environment is equal to $100 \times 100$ and 100 nodes.

Table 6. The impact of changing the speed of sink on network lifetime.

| Sink speed m/round | | | | | | |
|---|---|---|---|---|---|---|
| 144 | | 96 | | 48 | | |
| The last node | The last node | The first nodes | The first nodes | The last node | The first nodes | Life Time ╲ Algorithm |
| 1380 | 875 | 1780 | 1140 | 2325 | 1495 | Propose2 |
| 1200 | 790 | 1600 | 1055 | 2145 | 1413 | Propose1 |
| 635 | 470 | 1035 | 734 | 1575 | 1088 | ELEACH |
| 465 | 365 | 865 | 633 | 1405 | 987 | LEACH |

According to Table (6)increase in sink speed will reduce the network lifetime, then sink speed is directly related to energy consumption in network sensor. Therefore, the increase in sink speed increases energy consumption of sensors and ultimately reduces the lifetime of the network.

## 4.CONCLUSION AND RECOMMENDATIONS

Problem of routing based on clustering in wireless sensor network aimed at reducing energy consumption and maintaining net coverage was investigated in the current study. To achieve this goal, we attempt to layered the environment based on the size of the network environment, so we can have a dynamic layering in terms of network environments size. This protocol is, in fact, a routing approach based on layering environment, where every independent layer starts clustering and sends information to the top layer to reach the sink, in fact, using layering and multi-hop routing algorithms, we will reduce energy loss during data transmission. Also in each layer in the protocol, the optimum number of nodes and summarization of the sizes of the cluster is done independently. Similarly, by moving a sink in each round when the sink is spinning the environment, layers varies according to different situations of the sink.

Based on the comparisons carried out in different modes, it can be said that increase in node sizes,the size of the environment, homogeneous or heterogeneous of environment has a tremendous impact on the lifetime of sensor networks. In a way that increases in the sizes of nodes will prolong the lifetime of the network. Increase in size of environment reduces the lifetime of the network and the heterogeneity of the environment will increase the lifetime of environment.

The most important innovation of this research, is a specific mechanism based on layered environments that reduce energy consumption in sensors which are further away from the sink which applying this mechanism had a significant effect in reduction of energy consumption and balance networks and increasing longevity and maintaining network coverage.The second innovation is a cluster topology control method so that the numbers of members of each cluster head are almost evenly distributed and also has a tremendous impact on energy consumption of cluster head. The other innovation in this research is providing a way to control the nodes in the first layer of coverage so that the network coverage maintains and power of nodes which hasn't chosen as gateway nodes is not lost. This has done with marinating of nodes.

As future research work that can be done to get the protocol presented in this study, the following suggestions are presented:

- Applying other useful parameters for the formation of clusters.
- Applying different structure for layering the environment that considers how large or small first layers it is, by taking the environment.
- Use of different criteria for selecting the cluster and improvement of the proposed cost function.
- Apply the appropriate parameters for the gateway nodes.
- Choose a useful mechanism for cluster heads to select the best cluster head of top layer to send information.

## References

[1]  L. Borges, F. Velez, and A. Lebres, "Survey on the Characterization and Classification of Wireless Sensor Networks Applications," *IEEE Communications Surveys & Tutorials*, vol. XX, no. X. pp. 1–1, 2014.

[2]  H. Asharioun, H. Asadollahi, T.-C. Wan, and N. Gharaei, "A Survey on Analytical Modeling and Mitigation Techniques for the Energy Hole Problem in Corona-Based Wireless Sensor Network," *Wirel. Pers. Commun.*, vol. 81, no. 1, pp. 161–187, 2015.

[3]  F. Shaukat, "A Survey on Testing Network Applications and Protocols," vol. 2, no. 02. pp. 316–325, 2015.

[4]  G. Sara and D. Sridharan, "Routing in mobile wireless sensor network: a survey," *Telecommun. Syst.*, vol. 57, no. 1, pp. 51–79, 2014.

[5]  R. C. Carrano, D. Passos, L. C. S. Magalhaes, and C. V. N. Albuquerque, "Survey and Taxonomy of Duty Cycling Mechanisms in Wireless Sensor Networks," *Communications Surveys & Tutorials, IEEE*, vol. 16, no. 1. pp. 181–194, 2014.

[6]  "Energy-efficient routing protocols in wireless sensor networks A survey 2013." .

[7]  G. Han, J. Jiang, L. Shu, J. Niu, and H.-C. Chao, "Management and applications of trust in Wireless Sensor Networks: A survey," *Journal of Computer and System Sciences*, vol. 1. pp. 1–16, 2013.

[8]  S. A. Sert, H. Bagci, and A. Yazici, "MOFCA: Multi-objective fuzzy clustering algorithm for wireless

sensor networks," *Appl. Soft Comput.*, vol. 30, no. 0, pp. 151–165, May 2015.

[9] D. V Jose and G. Sadashivappa, "Mobile Sink Assisted Energy Efficient Routing Algorithm for Wireless Sensor Networks," *The World of Computer Science and Information Technology*, vol. 5, no. 2. pp. 16–22, 2015.

[10] A. Ali Ahmed, "An enhanced real-time routing protocol with load distribution for mobile wireless sensor networks," *Comput. Networks*, vol. 57, no. 6, pp. 1459–1473, Apr. 2013.

[11] W. R. Heinzelman, A. Chandrakasan, and H. Balakrishnan, "Energy-efficient communication protocol for wireless microsensor networks," *System Sciences, 2000. Proceedings of the 33rd Annual Hawaii International Conference on*. p. 10 pp. vol.2, 2000.

[12] S. Lindsey and C. S. Raghavendra, "PEGASIS: Power-efficient gathering in sensor information systems," in *IEEE Aerospace Conference Proceedings*, 2002, vol. 3, pp. 1125–1130.

[13] M. F. K. Abad and M. A. J. Jamali, "Modify LEACH Algorithm for Wireless Sensor Network," *Int. J. Comput. Sci. Issues*, vol. 8, no. 5, pp. 219–224, 2011.

[14] D. Mahmood, N. Javaid, S. Mahmood, S. Qureshi, A. M. Memon, and T. Zaman, "MODLEACH: A variant of LEACH for WSNs," in *Proceedings - 2013 8th International Conference on Broadband, Wireless Computing, Communication and Applications, BWCCA 2013*, 2013, pp. 158–163.

[15] M. Ghiasabadi, M. Sharifi, N. Osati, S. Beheshti, and M. Sharifnejad, "TEEN: a routing protocol for enhanced efficiency in wireless sensor networks," *2008 Second Int. Conf. Futur. Gener. Commun. Netw.*, vol. 1, no. C, pp. 2009–2015, 2001.

[16] A. Manjeshwar and D. P. Agrawal, "APTEEN: a hybrid protocol for efficient routing and comprehensive information retrieval in wireless," in *Proceedings 16th International Parallel and Distributed Processing Symposium*, 2002, p. 8 pp.

[17] O. Younis and S. Fahmy, "HEED: a hybrid, energy-efficient, distributed clustering approach for ad hoc sensor networks," *IEEE Trans. Mob. Comput.*, vol. 3, no. 4, pp. 366–379, 2004.

[18] S. Chand, S. Singh, and B. Kumar, "Heterogeneous HEED Protocol for Wireless Sensor Networks," *Wirel. Pers. Commun.*, vol. 77, no. 3, pp. 2117–2139, 2014.

[19] O. Boyinbode, H. Le, and M. Takizawa, "A survey on clustering algorithms for wireless sensor networks," *Int. J. Space-Based Situated Comput.*, vol. 1, no. 2–3, pp. 130–136, 2011.

[20] N. A. Pantazis, S. A. Nikolidakis, and D. D. Vergados, "Energy-Efficient Routing Protocols in Wireless Sensor Networks: A Survey," *IEEE Commun. Surv. Tutorials*, vol. 15, no. 2, pp. 551–591, 2013.

[21] K. Chen, "Unequal Cluster-Based Routing Protocol in Wireless Sensor Networks," *Journal of Networks*, vol. 8, no. 11. 2013.

[22] Y. Jiang, W. Shi, X. Wang, and H. Li, "A distributed routing for wireless sensor networks with mobile sink based on the greedy embedding," *Ad Hoc Networks*, vol. 20. pp. 150–162, 2014.

[23] Z. Han, J. Wu, J. Zhang, L. Liu, and K. Tian, "A general self-organized tree-based energy-balance routing protocol for wireless sensor network," *IEEE Transactions on Nuclear Science*, vol. 61, no. 2. pp. 732–740, 2014.

[24] J. Zhu, C.-H. Lung, and V. Srivastava, "A hybrid clustering technique using quantitative and qualitative data for wireless sensor networks," *Ad Hoc Networks*, vol. 25, Part A, no. 0, pp. 38–53, Feb. 2015.

## Authors

**Ali Sedighimanesh** is a graduate student in the School of Electrical and Computer Engineering, University of Science and Technology, Qazvin Islamic azad University (QIAU),Iran. he received a Bachelor degree from University of Science and Technology Parand, iran. His research areas are wireless communications and Network. and a Master degree from QIAU. His current research interests include wireless and mobile communications, cooperative communications, optimization theory on communications.

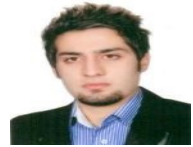

**Mohammad Sedighimanesh** is a graduate student in the School of Electrical and Computer Engineering, University of Science and Technology, Qazvin Islamic azad University(QIAU),Iran. he received a Bachelor degree from University of Science and Technology Zanjan, iran. His research areas are wireless communications and Network. and a Master degree from QIAU. His current research interests include wireless and mobile communications, cooperative communications, optimization theory on communications.

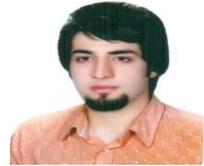

**Javad Baqeri** is a graduate student in the School of Electrical and Computer Engineering, University of Science and Technology, Qazvin Islamic azad University(QIAU),Iran. he received a Bachelor degree from University of Science and Technology Shomal, iran. His research areas are wireless communications and Network. and a Master degree from QIAU. His current research interests include wireless and mobile communications, cooperative communications, optimization theory on communications.

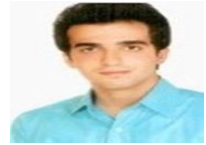